\documentclass[twocolumn,showpacs,floats,floatfix,superscriptaddress,aps,pra]{revtex4-1}
\usepackage{amsfonts}
\usepackage{amssymb}
\usepackage{amsmath}
\usepackage{calc}
\usepackage{graphicx}
\usepackage{bm}

\usepackage[normalem]{ulem}
\usepackage{amsmath,amssymb}
\usepackage{multirow}
\usepackage{xcolor,soul}
\usepackage{srcltx}
\usepackage{hyperref,graphicx}

\def\be{ \begin{equation}}
\def\ee{ \end{equation}}
\def\bea{ \begin{eqnarray}}
\def\eea{ \end{eqnarray}}
\def\bse{ \begin{subequations}}
\def\ese{ \end{subequations}}
\def\bc{ \begin{center}}
\def\ec{ \end{center}}

\begin{document}

\author{Stefano Longhi$^{*}$} 
\affiliation{Dipartimento di Fisica, Politecnico di Milano and Istituto di Fotonica e Nanotecnologie del Consiglio Nazionale delle Ricerche, Piazza L. da Vinci 32, I-20133 Milano, Italy}
\email{stefano.longhi@polimi.it}

\title{Metal-insulator phase transition in a non-Hermitian Aubry-Andr\'e-Harper Model}
  \normalsize


%
\bigskip
\begin{abstract}
\noindent  
Non-Hermitian extensions of the Anderson and Aubry-Andr\'e-Harper models are attracting a considerable interest as platforms to study localization phenomena, metal-insulator and topological phase transitions in disordered non-Hermitian systems. Most of available studies,  however, resort to numerical results, while few analytical and rigorous results are available owing to the extraordinary complexity of the underlying problem. Here we consider a parity-time ($\mathcal{PT}$)  symmetric extension of the Aubry-Andr\'e-Harper model, undergoing a topological metal-insulator phase transition, and provide rigorous analytical results of  energy spectrum, symmetry breaking phase transition and localization length. In particular, by extending to the non-Hermitian realm the Thouless$^{\prime}$s result relating localization length and density of states, we derive an analytical form of the localization length in the insulating phase, showing that -- like in the Hermitian Aubry-Andr\'e-Harper model-- the localization length is independent of energy. 
\end{abstract}

\maketitle
      
\section{Introduction}
The Aubry-Andr\'e-Harper (AAH) model \cite{R1,R2} provides a paradigmatic example of one-dimensional (1D) quasicrystal which has attracted a continuous interest both theoretically and experimentally since the past three decades (see, for instance, \cite{R3,R4,R5,R6,R7,R8,R9,R10,R11,R12,R13,R14,R15,R16,R17,R17bis} and references therein). 
The most peculiar feature of the AAH model is that, for a truly  incommensurate potential \cite{R18}, the system undergoes a sudden metal-insulator phase transition
at a finite critical value of the quasiperiodic potential strength with all localized eigenstates having the same localization length, which is a distinctive feature as compared to Anderson localization in disordered 1D lattices. At the critical point, the energy spectrum is governed by the Harper equation, which describes the motion of a quantum particle on a 2D crystal subjected to a magnetic flux and shows the characteristic Hofstadter butterfly energy spectrum \cite{R19}.\\
Recently, great interest has been devoted to study metal-insulator phase transitions and localization phenomena in non-Hermitian systems
\cite{R20,R21,R22,R23,R24,R25,R26,R27,R28,R29,R30,R31,R32,R33,R34,R35,R36,R37,R38,R39,R40,R41,R42,R43,R44,R45,R46,R47,R48,R49}, highlighting similarities and differences as compared to
ordinary Hermitian models. Several non-Hermitian models with  disorder or incommensurate potentials have been investigated, in which non-Hermiticity is introduced by considering either complex on-site potentials, phenomenologically describing dissipation and/or amplification with the surrounding enviroment, or asymmetric hopping amplitudes, such as in systems with synthetic imaginary gauge fields.
In systems with on-site  complex disorder (non-Hermitian Anderson model), it was shown that a purely imaginary disorder can induce localization like in the Hermitian Anderson localization problem, with a duality between dissipation and amplification \cite{R20,R21,R26,R29,R33}. On the other hand, for real-energy on-site potential disorder a non-Hermitian delocalization transition is observed by application of an imaginary gauge field (Hatano-Nelson-Anderson model \cite{R22,R23,R24,R25,R28,R32,R38,R45,R46,R48}). 
Other studies focused on several non-Hermitian extensions of  diagonal or off-diagonal AAH models \cite{R30,R34,R35,R36,R39,R40,R42,R43,R46,R47,R48,R49}, with either commensurate or incommensurate potential, showing the impact of non-Hermiticity terms in the Hamiltonian on edge states and parity-time ($\mathcal{PT}$) symmetry breaking \cite{R34,R35,R36,R40,R42}, on the Hofstadter butterfly spectrum \cite{R35},  and on the localization properties of eigenstates \cite{R42,R43,R46,R47,R48,R49}. Recently, the topological nature of a metal-insulator phase transition found in an incommensurate $\mathcal{PT}$-symmetric AAH model has been revealed \cite{R47}.\\ 
Most of available  results on localization and phase transition phenomena in non-Hermitian AAH  models are based on numerical simulations, while few analytical and rigorous results are available owing to the extraordinary complexity of the spectral problem. In particular, as compared to the Hermitian AAH model, in non-Hermitian AAH models the self-duality property is generally lost, and the complex nature of the energy spectrum makes it not straightforward to relate density of states and localization length. In this article we consider a $\mathcal{PT}$ symmetric extension of the Aubry-Andr\'e-Harper model, which is known to undergo a topological metal-insulator phase transition \cite{R30,R47}, and provide analytical results of  energy spectrum, symmetry breaking phase transition and localization length. In particular, by extending to the non-Hermitian realm the Thouless$^{\prime}$s result relating localization length and density of states \cite{R50} , we derive an analytical form of the localization length in the insulating phase, showing that -- like in the Hermitian AAH model-- it is independent of energy. On the other hand, unlike the Hermitian AAH model in the metallic phase the energy spectrum is gapless.

\section{Non-Hermitian Aubry-Andr\'e-Harper model}
\subsection{Model}
We consider a non-Hermitian $\mathcal{PT}$-symmetric extension of the AAH model with complex incommensurate on-site potential on a 1D lattice, described by the Hamiltonian
\begin{equation}
\hat{H} \psi_n=J(\psi_{n+1}+\psi_{n-1})+V_n \psi_n 
\end{equation}
for the occupation amplitudes $\psi_n$ at the various sites of the lattice, where $J$ is the hopping rate,  
\begin{equation}
V_n= V_0 \exp(-2 \pi i \alpha n)  
\end{equation}
is the onsite complex potential of amplitude $V_0$, and $\alpha$ is irrational for incommensurate potentials. As shown in \cite{R47}, such a non-Hermitian AAH model can be obtained from the ordinary
AAH model with on-site potential $V_n=A  \cos( 2 \pi \alpha n + \varphi)$ after complexification of the phase $\varphi=i h$ and taking the limits $A \rightarrow 0$, $h \rightarrow \infty$, with $A \exp(h)= 2 V_0$ finite. 
The eigenvalue equation for the Hamiltonian (1) reads
\begin{equation}
E \psi_n=J(\psi_{n+1}+\psi_{n-1})+V_0 \exp(- 2 \pi i \alpha n) \psi_n.
\end{equation}
Numerical results show that a metal-insulator phase transition arises at the critical point $V_0=J$ \cite{R30,R47}, which is signaled by a $\mathcal{PT}$ symmetric breaking phase transition of energy spectrum: for $V_0<J$ the energy spectrum is entirely real and all eigenstates are delocalized (metallic and unbroken $\mathcal{PT}$ phases), while for $V_0>J$ the energy spectrum becomes complex and all eigenstates are localized (insulating and $\mathcal{PT}$ broken phases); see Figs.1 and 2 for typical numerical results.
The phase transition is of topological nature and can be expressed in terms of a winding number \cite{R47}.\\
Let us briefly remind the main results of the localization-delocalization phase transition in the ordinary (Hermitian) AAH model, corresponding to the real potential $V_n=2 V_0 \cos( 2 \pi \alpha n)$ with $\alpha$ irrational. In this case, it is well known that the model
exhibits a transition from a metallic phase for $V_0< J$ to an insulating phase when $V_0>J$. The energy spectrum has a Cantor set structure. In the extend phase $\hat{H}$ has an absolutely
continuous gapped spectrum with a Lebesgue measure $4(J-V_0)$, which vanishes as the transition point $V_0=J$ is attained. In the insulating phase, all eigenstates are exponentially localized with the same Lyapunov exponent $\lambda=\log (V_0/J)$ independent of energy.
\begin{figure}[htbp]
  \includegraphics[width=86mm]{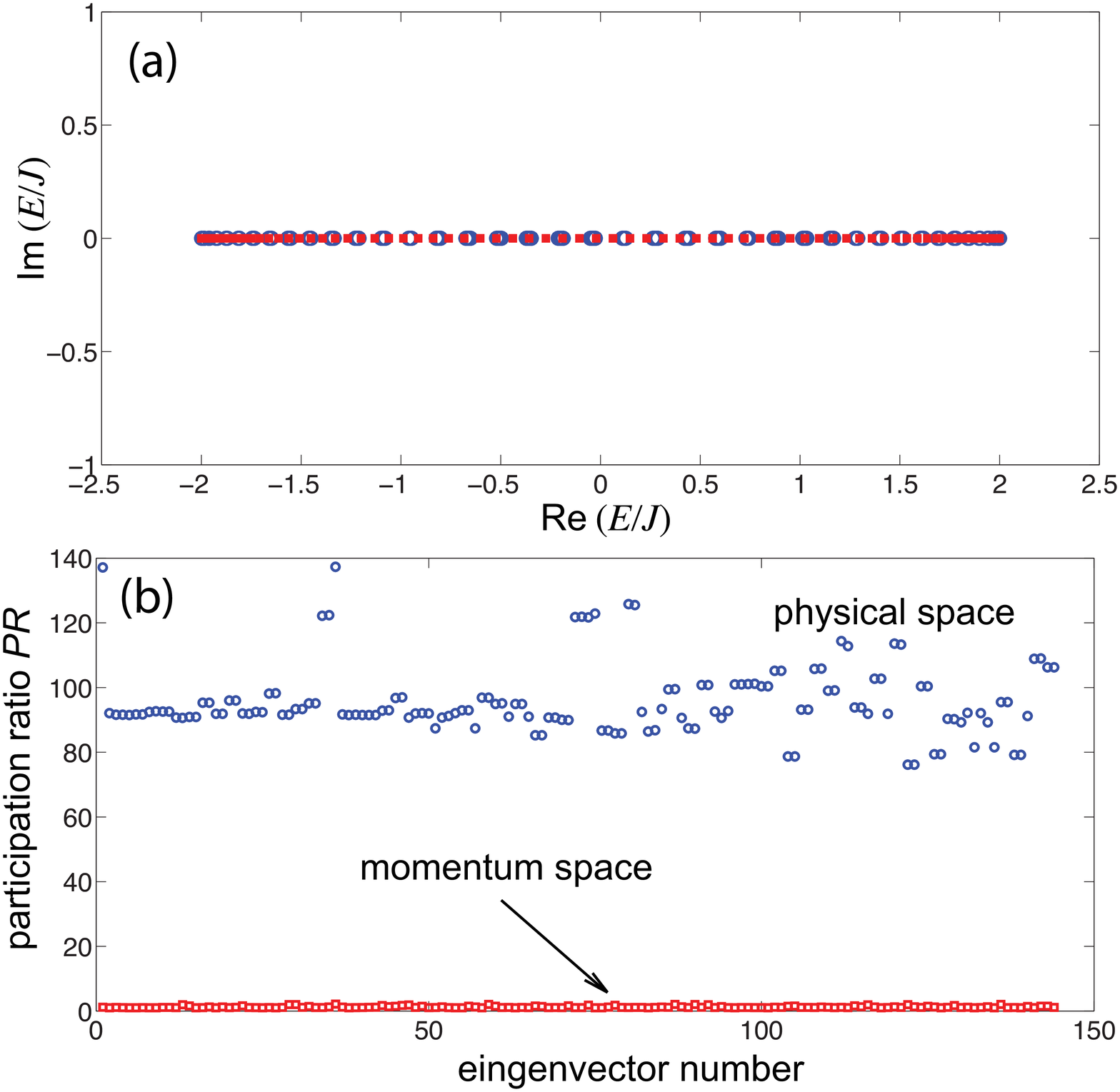}\\
   \caption{(color online) (a) Example of numerically-computed energy spectrum $E$ of the non-Hermitian AAH Hamiltonian $\hat{H}$ [Eq.(3)] in the metallic and unbroken $\mathcal{PT}$ phases ($V_0<J)$. Parameter values used in simulations are $\alpha=(\sqrt{5}-1)/2$ (the inverse of the golden mean) and $V_0/J=0.5$. The  approximant $\alpha \simeq p/q$ of the inverse of the golden mean, with $p=89$ and $q=144$, has been assumed. The lattice comprises $L=q=144$ sites and PBC are used to compute the energy spectrum. Open circles refer to the numerically-computed eigenvalues of $\hat{H}$, i.e. the roots of the characteristic polynomial $P(E)$ defined by Eq.(8), whereas  squares are defined by the relation $E=2 J \cos (k_l)$ with quantized wave number $k_l=2 l \pi/L$ ($l=1,2,3,...,L$).
 (b) Numerically-computed participation ratio $PR$ of the $L=144$ eigenvectors of the Hamiltonian $\hat{H}$  in real space [Eq.(3)] and of the isospectral Hamiltonian $\hat{H}_1$ in momentum space  [Eqs.(5) and (6)]. 
The participation ratio $PR$ is defined as $PR=( \sum_n |\psi_n|^2)^2 / \sum_n |\psi_n|^4$ for the eigenvectors $\psi_n$ of $\hat{H}$, and $PR=( \sum_n |\phi_n|^2)^2 / \sum_n |\phi_n|^4$  for the eigenvectors $\phi_n$ of $\hat{H}_1$. A value of $PR$ close to one indicates a fully localized eigenstate, wheres a large value of $PR$, of order $PR \sim L$, indicates a fully delocalized eigenstate. Note that for $V_0<J$ all eigenstates are delocalized in physical space (metallic phase) and localized in momentum space.}
\end{figure}
\begin{figure}[htbp]
  \includegraphics[width=86mm]{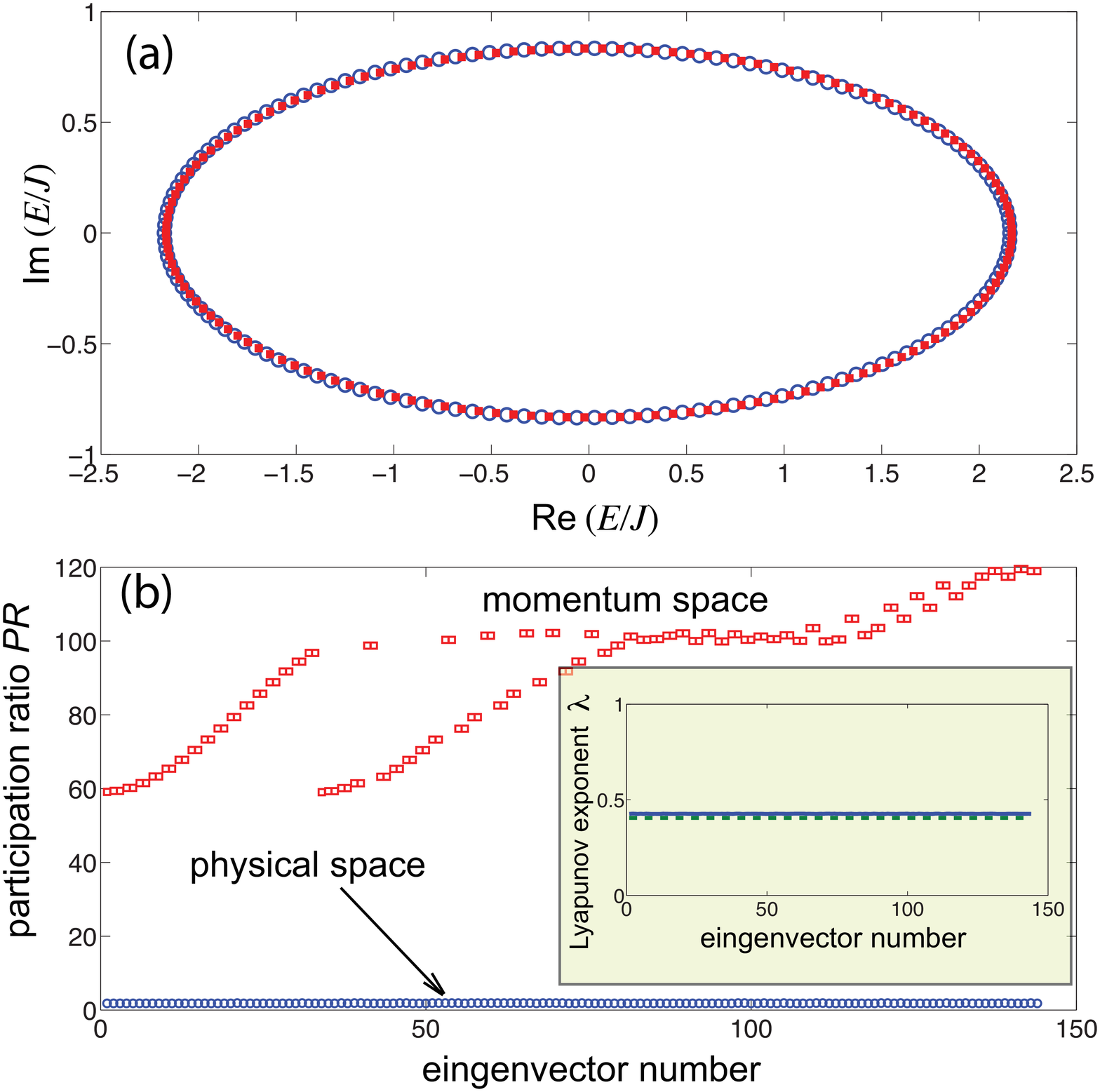}\\
   \caption{(color online) (a) Example of numerically-computed energy spectrum $E$ of the non-Hermitian AAH Hamiltonian $\hat{H}$ in the insulating  and broken $\mathcal{PT}$ phases ($V_0>J)$. Parameter values used in simulations are $\alpha=(\sqrt{5}-1)/2$ and $V_0/J=1.5$. Open circles refer to the numerically-computed eigenvalues of the $\hat{H}$, i.e. the roots of the polynomial $P(E)$ defined by Eq.(8), whereas  squares correspond to the points on the ellipse defined by the equation $E=2 J \cos (k_l-ih)$ [Eq.(15)] with $h= \log (V_0 /J)$ and with quantized wave number $k_l=2 l \pi/L$ ($l=1,2,3,...,L$).
 (b) Numerically-computed participation ratio $PR$ of the $L=144$ eigenvectors of the Hamiltonian $\hat{H}$  in real space [Eq.(3)] and of the isospectral Hamiltonian $\hat{H}_1$ in momentum space  [Eqs.(5) and (6)]. 
Note that for $V_0>J$ all eigenstates are localized in physical space (insulating phase) and delocalized in momentum space. The inset in (b) shows the numerically-computed Lyapunov exponent $\lambda$ of the various eigenvectors (solid curve) and the theoretical value given by Eq.(33) (dashed curve).}
\end{figure}
\subsection{Spectral problem in momentum space}
As is well-known, the ordinary (Hermitian) AAH model with nearest-neighbor hopping is self-dual, i.e. the spectral problem in momentum and real spaces are described by the same Hamiltonian with hopping and potential amplitudes interchanged \cite{R3}.  The self-dual property of the Hamiltonian is extremely useful to provide analytical insights into the metal-insulator phase transition and to calculate the localization length of eigenstates in the insulating phase. Unfortunately, as shown below in the non-Hermitian AAH model with complex on-site potential, defined by Eq.(3), the self-duality is broken, and so far there are not analytical results regarding the localization length and complex energy spectrum in the insulating (broken $\mathcal{PT}$) phase.\\
To introduce the spectral problem in momentum space, let us consider Eq.(3) in real space on a ring, comprising $L$ sites, with periodic boundary conditions (PBC) $\psi_{n+L}=\psi_n$, and let us then consider the thermodynamic limit $L \rightarrow \infty$. More precisely, we consider an approximant of the irrational number $\alpha$ defined by a sequence $p_n/q_n$ of rational and irreducible numbers ($n=1,2,3,...$), with $q_n,p_n \rightarrow \infty$ as $n \rightarrow \infty$. Then, we assume $L=q_n$ and take the limit $n \rightarrow \infty$. Note  that, since the non-Hermitian Hamiltonian $\hat{H}$ does not show the non-Hermitian skin effect \cite{R45,R47,R51,R52,R53,R54,R55}, the bulk energy spectrum of $\hat{H}$ is insensitive to the boundary conditions, i.e. the energy spectrum and localization properties of bulk states do not change when considering open boundary conditions (OBC) rather than PBC. The problem in momentum space is formulated by introduction of the discrete Fourier transform
\begin{equation}
\phi_n= \frac{1}{\sqrt{L}} \sum_{l=1}^L \psi_l \exp(2 \pi i \alpha l n)
\end{equation}
i.e. 
\begin{equation}
\psi_n= \frac{1}{\sqrt{L}} \sum_{l=1}^L \phi_l \exp(-2 \pi i \alpha l n).
\end{equation}
The eigenvalue equation (3) is then transformed into the following one
\begin{equation}
E \phi_n=V_0  \phi_{n-1} +W_n \phi_n \equiv \hat{H}_1 \phi_n
\end{equation}
where we have set
\begin{equation}
W_n= 2 J \cos (2 \pi \alpha n)
\end{equation}
and where the periodic boundary conditions $\phi_{n+L}=\phi_n$ hold. Note that $\hat{H}$ and $\hat{H}_1$ are related one another by a similarity transformation \cite{R47} and thus, under PBC, they are isospectral. Clearly, if $\psi_n$ is a localized (proper) eigenstate of $\hat{H}$ in real space, then $\phi_n$, defined by Eq.(4),  is a delocalized (improper) eigenstates of $\hat{H}_1$ in momentum space, and vice-versa. Unlike the Hermitian AAH model, a comparison of Eqs.(3) and (6) indicates that the non-Hermitian AAH model is not self-dual. However, under PBC the Hamiltonians $\hat{H}$ and $\hat{H}_1$ are isospectral, with localization/delocalization nature of eigenstates interchanged. 

\section{Metal-insulator phase transition and localization length}
In some recent works \cite{R30,R47}, it has been shown by extended numerical analysis that the metal-insulator transition of the Hamiltonian (3), from an extended (metallic) to a localized (insulating) phase as $V_0$ is increased above $J$, corresponds to a $\mathcal{PT}$ symmetry breaking phase transition, i.e. to a transition of the energy spectrum from being entirely real for $V_0<J$ to complex energies for $V_0>J$ (Figs.1 and 2). 
The bulk energy spectrum of $\hat{H}$ and $\hat{H}_1$, under PBC, are the same and can be obtained from the roots $E_{n}$ of the following characteristic polynomial $P(E)$ of order $L$, as shown in the Appendix A
\begin{equation}
P(E)=\prod_{l=1}^L(E-W_l)-V_0^L
\end{equation}
with $W_l=2 J \cos (2 \pi \alpha l)$. An extended numerical analysis of the roots of polynomial $P(E)$ shows that, for $V_0<J$ and in the large $L$ limit, the roots are real and fill the interval $( -2J,2J)$ with a density of states $\rho(E) \sim L/ \sqrt{4J^2- E^2}$ [Fig.1(a)], while for $V_0>J$ they become complex and lie of an ellipse of the complex energy plane [Fig.2(a)].
Here we derive analytical expression of the energy spectrum in both phases and of the localization length of eigenstates in the insulating phase, which avoids the numerical computation of polynomial roots.\\
\subsection{Extended phase}
Let us consider the case $V_0<J$.  In this regime numerical results show that all eigenstates of $\hat{H}$ in real space are extended and the energy spectrum is entirely real (unbroken $\mathcal{PT}$ phase), filling the interval $(-2J,2J)$\cite{R30,R47}. Here we provide a rigorous proof of such results. Let us consider the Hamiltonian $\hat{H}_1$ in momentum space [Eq.(6)]. For an arbitrary integer $n_0$, a formal solution to Eq.(6) is given by
\begin{equation}
\phi_n \propto \left\{
\begin{array}{cc}
0 & n <n_0 \\
1  & n=n_0 \\
\frac{V_0}{E-W_n} \phi_{n-1} & n >n_0
\end{array}
\right.
\end{equation}
with energy $E=W_{n_0}=2J \cos(2 \pi \alpha n_0)$. Such a solution is an eigenstate of $\hat{H}_1$ provided that $|\phi_n|$ does not diverge as $n \rightarrow \infty$. In particular, if $\phi_n \rightarrow 0$ as $n \rightarrow \infty$ and $\sum_n |\phi_n|^2 < \infty$, the eigenstate is localized. The localization properties of the solution (9) are derived by computing the Lyapunov exponent
\begin{widetext}
\begin{equation}
\mu(E) =  - \lim_{n \rightarrow \infty} \frac{1}{n-n_0} \log \left| \frac{\phi_{n}}{\phi_{n_0}} \right| = \lim_{n \rightarrow \infty} \frac{1}{n-n_0} \sum_{k=n_0+1}^{n}  \log \left|  \frac{2J \cos(2 \pi \alpha k)-2J \cos (2 \pi \alpha n_0)}{V_0} \right| 
\end{equation}
\end{widetext}
with the requirement $\mu>0$ for localization.
Let $q_0=2 \pi \alpha n_0$ mod. $2 \pi$, and let us observe that, since $\alpha$ is irrational, $q=2 \pi \alpha k$ mod. $ 2 \pi$ uniformly fills the interval $(-\pi,\pi)$ as $k$ varies from $n_0+1$ to $\infty$. 
This follows from the Weyl$^{\prime}$s equidistribution theorem and properties of irrational rotations, which are dense  in the interval $(-\pi,\pi)$ and ergodic with respect to the Lebesgue measure \cite{R56,R57}.
Hence in the large $n$ limit Eq.(10) takes the form
\begin{equation}
\mu(E)= \frac{1}{2 \pi}  \int_{-\pi}^{\pi} dq \log \left( \frac{2J}{V_0} \left|  \cos(q)-\cos(q_0)\right|  \right)  
\end{equation}
i.e.
\begin{equation}
\mu(E)= \log \left( \frac{J}{V_0} \right) + \log 2 + \frac{1}{2 \pi}  \int_{-\pi}^{\pi} dq \log \left|  \cos(q)-\cos(q_0)\right| . 
\end{equation}
The integral on the right hand side of Eq.(12) is independent of $q_0$  (see Appendix B for technical details) and can be evaluated for $q_0=\pi/2$. Taking into account that
\begin{equation}
\int_{-\pi}^{\pi} dq \log \left|  \cos(q) \right|=-2 \pi \log 2
\end{equation}
one finally obtains
\begin{equation}
\mu(E)= \log \left( \frac{J}{V_0} \right) . 
\end{equation}
for the Lyapunov exponent. Note that $\mu(E)$ is independent of the eigenenergy $E$. Localization in momentum space ($\mu>0$) requires $V_0<J$, corresponding to extended eigenstates $\psi_n$ of $\hat{H}$ in real space. This result demonstrates that for $V_0 <J$ the system in real space is in the metallic phase and that the energy spectrum is absolutely continuous and entirely real, describing the gapless interval  $-2J \leq E \leq 2J$. Moreover, since $q_0$ uniformly fills the interval $(- \pi, \pi)$ as the integer $n_0$ is varied, the energy spectrum can be written as $E=2J \cos(q_0)$, resulting in a density of states $\rho(E)=(L/ 2 \pi) |dE/dq_0|^{-1}=  (L/ 2 \pi)(4J^2-E^2)^{-1/2}$. This means that, rather counterintuitively, the energy spectrum and density of states of $\hat{H}$ in the metallic phase are independent of the potential strength $V_0$ and thus are the same as the one of the potential-free Hamiltonian $V_0=0$. Remarkably, in contrast to  the ordinary (Hermitian) AAH model [Eq.(1) with $V_n=2 V_0 \cos (2 \pi \alpha n)$], whose spectrum is gapped, in the non-Hermitian AAH model (3) the eigenspectrum is gapless.

 \subsection{Localized phase}
 {\it Energy spectrum.}\\
 \\
For $V_0>J$, all eigenstates of $\hat{H}$ in real space are exponentially localized and the energy spectrum becomes complex (broken $\mathcal{PT}$ phase). Extended numerical results lead to conjecture \cite{R30}  the following 
expression for the eigenenergies $E$ 
\begin{eqnarray}
E & = & \left( V_0+\frac{J^2}{V_0} \right) \cos(k)+i \left( V_0-\frac{J^2}{V_0} \right) \sin(k) \nonumber \\
& = & 2 J \cos(k-ih)
\end{eqnarray}
with $ -\pi \leq k < \pi$ and $h \equiv \log(V_0/J)$, i.e.
\begin{equation}
\left( \frac{{\rm Re}(E)}{V_0+J^2/V_0} \right)^2+\left( \frac{{\rm Im}(E)}{V_0-J^2/V_0} \right)^2=1
\end{equation}
corresponding to an ellipse in the complex energy plane [Fig.2(a)]. Here we provide a rigorous proof of Eq.(15) and, most important, we derive an analytical expression of the localization length of the eigenstates, showing that --like in the Hermitian AAH model-- it is independent of energy $E$. To this aim, let us first consider the dual Hamiltonian $\hat{H}_1$ in momentum space and let us prove that, whenever the energy $E$ in Eq.(6) is chosen according to Eq.(15), the solution $\phi_n$ to Eq.(6) is an improper (non-normalizable) eigenfunction of $\hat{H}_1$, i.e. the Lyapunov exponent
\begin{equation}
\mu (E) =  - \lim_{n \rightarrow \infty} \frac{1}{n} \log \left| \frac{\phi_{n}}{\phi_{0}} \right| 
\end{equation}
vanishes. This means that $E$ belongs to the point spectrum of $\hat{H}$ with normalizable (localized) eigenfunctions $\psi_n$. To calculate the Lyapunov exponent $\mu$ in momentum space, we note that from Eq.(6) one has 
\begin{equation}
\frac{\phi_n}{\phi_{n-1}}= \frac{V_0}{E-2 J \cos (2 \pi \alpha n)},
\end{equation}
i.e. 
\begin{equation}
\frac{\phi_n}{\phi_0}= \prod_{k=1}^{n} \frac{\phi_k}{\phi_{k-1}}= \prod_{k=1}^{n} \frac{V_0}{E-2 J \cos (2 \pi \alpha k)}.
\end{equation}
Substitution of Eq.(19) into Eq.(17) yields
\begin{equation}
\mu(E)=\lim_{n \rightarrow \infty} \frac{1}{n} \sum_{k=1}^{n} \log \left| \frac{E-2J \cos(2 \pi \alpha k)}{V_0} \right| .
\end{equation}
For irrational $\alpha$, we can use again the Weyl$^{\prime}$s equidistribution theorem and ergodic property of irrational rotations, obtaining
\begin{equation}
\mu(E)=\frac{1}{2 \pi} \int_{-\pi}^{\pi} dq \log \left| \frac{E-2J \cos(q)}{V_0} \right|,
\end{equation}
i.e.
\begin{equation}
\mu(E)=-\log V_0+{\rm Re}( Q(E))
\end{equation}
where we have set
\begin{equation}
Q(E)=\frac{1}{2 \pi} \int_{-\pi}^{\pi} dq \log \left( E-2J \cos(q) \right).
\end{equation}
The integral on the right hand side of Eq.(23) can be analytically computed and reads (see Appendix B for technical details) 
\begin{equation}
Q(E)=i \theta+ \log J
\end{equation}
where the complex angle $\theta$ is defined by the relation
\begin{equation}
\cos \theta \equiv \frac{E}{2J}
\end{equation}
with ${\rm Im}(\theta)<0$.
 From Eqs.(22) and (24) one obtains
 \begin{equation}
 \mu(E)=-\log \left( \frac{V_0}{J} \right)- {\rm Im}(\theta)=-h- {\rm Im}(\theta).
 \end{equation}
 The energy spectrum in the localized phase is obtained by letting $\mu(E)=0$, corresponding to  $\theta=k-ih$ with $k$ an arbitrary real number. Then from Eq.(25) one obtains $E=2 J \cos(k-ih)$, which describes an ellipse as $k$ spans the range $(-\pi, \pi)$, according to the conjecture of Eq.(16).\\
\\
{\it Localization length.}\\
\\
To compute the localization length $\xi$ of the eigenstate $\psi_n$ with eigenenergy $E=E_n$ in real space, we calculate the Lyapunov exponent 
\begin{equation}
\lambda(E)= \lim_{l \rightarrow \infty} \left|  \frac{1}{l} \log \left|  \frac{\psi_l}{\psi_0} \right|\right|
\end{equation}
which is the inverse of the localization length $\xi(E)$. In Hermitian tight-binding models with nearest-neighbor hopping, Thouless developed a rather general relation that gives the Lyapunov exponent in terms of an integral involving the density of states \cite{R50,R59}. In non-Hermitian tight-binding lattices with nearest-neighbor hopping, we can generalize Thouless$^{\prime}$ result and show that, provided that the hopping amplitudes are symmetric (i.e. in the absence of the non-Hermitian skin effect), a similar relation can be established. For homogeneous and symmetric hopping amplitude $J$ (like in our non-Hermitian AAH model), one obtains (see Appendix C)
\begin{equation}
\lambda(E)=\lim_{L \rightarrow \infty} \frac{1}{L}\sum_{l=1, \; l \neq n}^{L} \log \frac{ \left| E_n-E_l \right|}{J}
\end{equation}
where $E_l$ ($l=1,2,3,...,L$) are the eigenvalues of $H$ for a chain of size $L$. 
Note that, after introduction of the density of states per unit length  $\rho(E^{\prime})$ in complex energy plane $E^{\prime}=E^{\prime}_R+iE^{\prime}_I$, such that $L \rho(E^{\prime})dE^{\prime}_RdE^{\prime}_I$ is the number of eigenstates of $\hat{H}$ with complex energy $E^{\prime}$ inside the infinitesimal square of vertices $(E^{\prime}_R,E^{\prime}_I)$, $(E^{\prime}_R+dE^{\prime}_R, E^{\prime}_I)$, $(E^{\prime}_R, E^{\prime}_I+d E^{\prime}_I)$, $(E^{\prime}_R+dE^{\prime}_R, E^{\prime}_I+d E^{\prime}_I)$, Eq.(28) can be written in the following form, which generalizes the well-know Thouless$^{\prime}$ result
\begin{equation}
\lambda(E)=  \int \int dE^{\prime}_R dE^{\prime}_I \rho(E^{\prime}) \log \frac{| E-E^{\prime} | }{J}.
\end{equation}
To calculate the localization length for the eigenstates of the non-Hermitian AAH Hamiltonian $\hat{H}$, let us make use of the following identity, which is proven in the Appendix A
\begin{equation}
\prod_{l=1 , \; l \neq n}^L \frac{E_{n}-E_{l}}{J}= \left( \frac{V_0}{J} \right)^L \sum_{l=1}^L \frac{J}{E_{n}-W_l}
\end{equation}
In the large $L$ limit, from Eqs.(28) and (30) one readily obtains
\begin{equation}
\lambda(E)=\log \left( \frac{V_0}{J} \right)
\end{equation}
since
\begin{eqnarray}
\lim_{L \rightarrow \infty} \frac{1}{L}  \log \left | \sum_{l=1}^L  \frac{J}{E-W_l}  \right|  \;\;\;\;\; \\
= \lim_{L \rightarrow \infty} \frac{1}{L} \log \left|  \frac{L}{2 \pi} \int_{-\pi}^{\pi} dk \frac{J}{E-2J \cos(k)} \right|=0. \nonumber
\end{eqnarray}
Hence the Lyapunov exponent (and hence the localization length $\xi(E)=1/ \lambda(E)$) is independent of energy $E$ and given by
\begin{equation}
\lambda(E)={\log \left( \frac{V_0}{J} \right)}.
\end{equation}
Such a theoretical result is in excellent agreement with the numerical simulations, as shown in the example of Fig.2(b).
\section{Conclusions}
Localization phenomena in non-Hermitian models, such as in non-Hermitian extensions of the Anderson and Aubry-Andr\'e-Harper models, are attracting a great interest since the past recent years, providing a platform to study localization phenomena, metal-insulator and topological phase transitions in disordered non-Hermitian systems. Most of available studies resort to numerical results, while there are very few analytical and rigorous results when dealing with non-Hermitian systems. In the Hermitian Aubry-Andr\'e-Harper model, analytical results are available owing to the self-dual property of the Hamiltonian and the ability to provide a simple analytical relation between localization length and density of states. Regrettably, in non-Hermitian extensions of the Aubry-Andr\'e-Harper model self-duality is generally broken and so far there have not been attempts to relate density of states and localization length in Hamiltonians with complex energies. In this work we filled such main gaps considering
a $\mathcal{PT}$  symmetric extension of the Aubry-Andr\'e-Harper model, which undergoes a topological metal-insulator phase transition. Such models could be physically implemented in photonic and electronic systems, as discussed in some recent works \cite{R46,R47,R48,R49,R60}. We  provided rigorous analytical results of  energy spectrum, symmetry breaking phase transition and localization length, confirming previous conjectures based on numerical results \cite{R30}. In particular, by extending to the non-Hermitian realm the Thouless$^{\prime}$s result relating localization length and density of states \cite{R50}, we derive an analytical form of the localization length in the insulating phase for the $\mathcal{PT}$  symmetric Aubry-Andr\'e-Harper model. Our results rigorously demonstrate the following similarities/differences between Hermitian and non-Hermitian Aubry-Andr\'e-Harper models:\\
(i) In the metallic (delocalized) phase, the energy spectrum of the non-Hermitian AAH model is entirely real (unbroken $\mathcal{PT}$ phase) and gapless.\\
(ii) In the metallic phase, the energy spectrum and density of states of the non-Hermitian AAH model are independent of the potential strength $V_0$ and irrational $\alpha$, and thus they are the same as the one of the potential-free Hamiltonian $V_0=0$. This is in stark contrast to the Hermitian AAH model, whose energy spectrum is gapped.\\
(iii) In the insulating (localized) phase, the energy spectrum of the AAH model is complex (broken $\mathcal{PT}$ phase) and describes an ellipse in complex energy plane. This means that all the roots of the characteristic polynomial associated to the non-Hermitian AAH model lie on an ellipse, a result that could be of relevance in number theory and polynomials \cite{R61,R62}.\\
(iv) For both Hermitian and non-Hermitian AAH models, in the insulating phase all eigenstates have the same localization length.\\

\appendix
\section{Energy spectrum and characteristic polynomial}
Under PBC, the Hamiltonians $\hat{H}$ and $\hat{H}_1$ in real and momentum space are isospectral, since they are obtained one another by a similarity transformation. The eigenvalues $E$ of the matrix $H_1$ [Eq.(6) in the main text] are obtained as the roots of the characteristic polynomial $P(E)=\det(E-H_1)$, which reads explicity
\begin{widetext}
\begin{equation}
 P(E)=
\left| 
\begin{array}{cccccccc}
E-W_1 & 0 & 0 & 0& ... & 0 & 0 & -V_0 \\
-V_0 & E-W_2 & 0 & 0 & ... & 0 & 0 & 0 \\
0 & -V_0 & E-W_3 & 0 & ... & 0 & 0 & 0 \\
... & ... & ... & ...& ... & ... & ... & ...  \\
0 & 0 & 0 & 0 & ... & -V_0 & E-W_{L-1} & 0 \\
0 & 0 & 0 & 0 & ... & 0 & -V_0 & E-W_L 
\end{array}
\right| 
\end{equation}
\end{widetext}
where $W_l=2 J \cos (2 \pi \alpha l$) $(l=1,2,...,L$). The determinant on the right hand side of Eq.(A1) can be readily computed from the first row, yielding
\begin{widetext}
\small
\begin{equation}
P(E)=(E-W_1)
\left| 
\begin{array}{ccccccc}
E-W_2 & 0 & 0 & ...  & 0 & 0 \\
 -V_0 & E-W_3 & 0 & ...  & 0 & 0 \\
... & ... & ...& ... & ... & ...  \\
 0 & 0 & 0 & ...  & E-W_{L-1} & 0 \\
 0 & 0 & 0 & ... & -V_0 & E-W_L 
\end{array}
\right| +(-1)^L V_0 
\left| 
\begin{array}{cccccccc}
-V_0 & E-W_2 & 0 &  ... & 0 & 0 \\
0 & -V_0 & E-W_3 &  ... & 0 & 0 \\
... & ... & ... & ...&  ... & ...  \\
0 & 0 & 0 &  ... & -V_0 & E-W_{L-1} \\
0 & 0 & 0 & ... & 0 &  -V_0 
\end{array}
\right|
\end{equation}
\end{widetext}
i.e.
\begin{equation}
P(E)=\prod_{l=1}^L (E-W_l)-V_0^L
\end{equation}
which is Eq.(8) given in the main text. It is worth showing the following interesting property of the eigenvalues $E_{\alpha}$ of $H_1$, i.e. of the roots of the characteristic polynomial $P(E)$
\begin{equation}
\prod_{\alpha=1 \; , \alpha \neq \beta}^{L} \frac{(E_{\beta}-E_{\alpha})}{J}=\left( \frac{V_0}{J} \right)^L \sum_{\alpha=1}^L \frac{J}{E_{\beta}-W_{\alpha}}
\end{equation}
In fact, let us calculate the derivative $P^{\prime}(E)$ of the characteristic polynomial. From Eq.(A3) one obtains
\begin{eqnarray}
P^{\prime}(E) & = & \sum_{\alpha=1}^L \prod_{n=1 \; , n \neq \alpha}^L (E-W_{n}) \nonumber \\
& = & \left( P(E)+V_0^L \right) \sum_{\alpha=1}^L\frac{1}{E-W_{\alpha}}
\end{eqnarray}
so that for $E=E_{\beta}$ one has
\begin{equation}
P^{\prime}(E_{\beta})=V_0^L \sum_{\alpha=1}^L \frac{1}{E_{\beta}-W_{\alpha}}.
\end{equation}
On the other hand, we can write
\begin{equation}
P(E)=\prod_{l=1}^L (E-E_l)
\end{equation}
so that
\begin{equation}
P^{\prime} (E)=\sum_{l=1}^L \prod_{\alpha=1 \; \alpha \neq l}^L (E-E_{\alpha})\end{equation}
and thus
\begin{equation}
P^{\prime} (E_{\beta})=\prod_{\alpha=1 \; \alpha \neq \beta}^L (E_{\beta}-E_{\alpha}).
\end{equation}
A comparison of Eqs.(A6) and (A9) yields
\begin{equation}
\prod_{\alpha=1 ,\; \alpha \neq \beta}^L (E_{\beta}-E_{\alpha})=V_0^L \sum_{\alpha=1}^L \frac{1}{E_{\beta}-W_{\alpha}}
\end{equation}
from which Eq.(A4) given above is readily obtained.
\section{Some useful integrals}
In the proofs given in the main text, we are required to calculate the following integral
\begin{equation}
Q(E)=\frac{1}{2 \pi} \int_{-\pi}^{\pi} dk \log \left( E-2 J \cos k \right)
\end{equation} 
with energy $E$ generally a complex, outside the interval $(-2J,2J)$ of the real energy axis. The definite integral given by Eq.(B1) is exactly solvable for real values of energy $E$; according to 
Gradshteyn and Ryzhik \cite{R58}, one has
\begin{equation}
Q(E)=\log \frac{E+\sqrt{E^2-4J^2}}{2}
\end{equation}
with $E$ real and $|E|>2 J$. We wish to extend the result (B.2) into the complex energy plane. To this aim, for any complex energy $E$ let us introduce the complex angle $\theta$ defined by
\begin{equation}
E=2 J \cos \theta
\end{equation}
with ${\rm Im}(\theta)<0$. Then
\begin{equation}
Q(\theta)=\frac{1}{2 \pi} \int_{-\pi}^{\pi} dk  \log \left( 2J \cos \theta- 2 J \cos k  \right)
\end{equation}
and thus, after taking the derivative with respect to $\theta$
\begin{equation}
\frac{dQ}{d\theta}=- \frac{\sin \theta}{2 \pi} \int_{-\pi}^{\pi} dk \frac{1}{\cos \theta- \cos k}.
\end{equation}
The integral on the right hand of Eq.(B5) can be computed after the substitution $z=\exp(ik)$ and using the residue theorem. In fact, after letting $z=\exp(ik)$, one has
\begin{eqnarray}
\int_{-\pi}^{\pi} dk \frac{1}{\cos \theta- \cos k} & = & 2 i \oint_{|z|=1} \frac{dz}{z^2-2 \cos \theta z+1} \nonumber \\
& = & 2 i \oint_{|z|=1} \frac{dz}{(z-z_1)(z-z_2) }
\end{eqnarray}
where the integral in the complex $z$ variable is extended over the unit circle $|z|=1$ and where we have set $z_1=\exp(i \theta)$, $z_2=\exp(-i \theta)$. Since ${\rm Im} (\theta)<0$, there is one pole, at $z=z_2$, inside the unit circle, while the other pole, at $z=z_1$, falls outside the unit circle and thus does not contribute to the integral. From the residue theorem one readily obtains
\begin{eqnarray}
\oint_{|z|=1} \frac{dz}{(z-z_1)(z-z_2) } & = & \frac{ 2 \pi i}{\exp(-i \theta)-\exp(i \theta)} \nonumber \\
& = &- \frac{\pi}{\sin \theta}
\end{eqnarray}
and thus, form Eqs.(B5), (B6) and (B7), one finally obtains
\begin{equation}
\frac{dQ}{d \theta}=i
\end{equation}
independent of $\theta$. After integration one obtains
\begin{equation}
Q(\theta)=i \theta+ {\rm const}.
\end{equation}
In can be readily shown that the integration constant on the right hand side of Eq.(B9) is equal to $\log J$. In fact, for $\theta=-i \psi$, with $\psi$ real and positive, the energy $E$ is real and given by $E=2J \cosh(\psi)$. In this limit Eq.(B9) should reduce to Eq.(B2). Taking into account that 
\begin{equation}
\frac{E+\sqrt{E^2-4J^2}}{2}=J \exp(i \theta)=J \exp( \psi)
\end{equation}
 one obtains ${\rm const}= \log J$, and thus
 \begin{equation}
 Q(E)=i \theta+\log J=i \arccos \left( \frac{E}{2J} \right)+\log J.
 \end{equation}
 Finally, let notice that the other integral
 \begin{equation}
 I(E) \equiv \frac{1}{2 \pi} \int_{-\pi}^{\pi} dk \log \left| E-2 J \cos(k) \right|
 \end{equation}
is simply obtained from $Q(E)$ using the relation 
\begin{equation}
I(E)={\rm Re} \left( Q(E) \right)=\log J- {\rm Im} (\theta).
\end{equation}
The limiting case of $E=2 J \cos (q_0)$ real inside the interval $(-J,J)$ can be obtained by letting $\theta=q_0-i \epsilon$, with $\epsilon>0$ and $\epsilon \rightarrow 0^+$. This yields
\begin{equation}
I(E=2J \cos q_0)=\log J
\end{equation} 
independent of $E$.
\section{Relation between localization length and density of states in a non-Hermitian tight-binding lattice}
In this Appendix we derive a simple relation between the localization length and the density of states in a non-Hermitian 1D tight-binding lattice with nearest-neighbor hopping, generalizing to the non-Hermitian realm the results derived by Herbert, Jones and Thouless in Refs. \cite{R50,R59}.  The eigenvalue equation of a single-band lattice with nearest-neighbor hopping described by the Hamiltonian $\hat{H}$  reads
\begin{equation}
\hat{H} \psi_n \equiv t_n \psi_{n+1}+\rho_{n-1} \psi_{n-1}+V_n \psi_n=E \psi_n
\end{equation}
where $t_n$, $\rho_{n-1}$ are the left/right hopping amplitudes between sites $n$ and $(n+1)$, and $V_n$ is the onsite potential. The Hermitian limit is obtained when $V_n=V^*_n$ and $\rho_n=t_n^*$. 
Non-Hermiticity is introduced by breaking either one of the two previous conditions. Provided that $|\rho_n|=|t_n|$, the non-Hermitian Hamiltonian $\hat{H}$ does not show the non-Hermitian skin effect, i.e. the squeezing of the eigenstates at either one of the two edges of the lattice for OBC, and the bulk energy spectrum in the thermodynamic limit does not depend on the specific boundary conditions.
To establish a relation between localization length and the density of states,  we assume a finite chain comprising $L$ sites with OBC, and then consider the thermodynamic limit $L \rightarrow \infty$.
Let us indicate by $\psi_n^{(\alpha)}$ the eigenvector of the matrix $H$
\begin{equation}
H= \left(
\begin{array}{cccccccccc}
V_1 & t_1 & 0 & 0 & 0 & ... & 0 & 0 & 0 & 0 \\
\rho_1 & V_2 & t_2 & 0 & 0 & ... & 0 & 0 & 0 & 0 \\
0 & \rho_2 & V_3 & t_3 & 0 & ... & 0 & 0 & 0 & 0 \\
0 & 0 & \rho_3 & V_4 & t_4 & ... & 0 & 0 & 0 & 0 \\
... & ... & ... & .. & ... & ... & .. & ... & ... & ... \\
0 & 0 & 0 & 0 & 0 & ... & \rho_{L-3} & V_{L-2} & t_{L-2} & 0 \\
0 & 0 & 0 & 0 & 0 & ... & 0 &  \rho_{L-2} & V_{L-1} & t_{L-1} \\
0 & 0 & 0 & 0 & 0 & ... & 0 &  0 &  \rho_{L-1} & V_{L}  \\
\end{array}
\right)
\end{equation}
corresponding to the eigenenergy $E_{\alpha}$ ($\alpha=1,2,3,...,L$). We assume that  $\psi_n^{(\alpha)}$ form a complete basis, i.e. that there are not exceptional points, corresponding to the coalescence of two or more eigenvectors of $H$. With  $\Psi_n^{(\alpha)}$ we indicate the eigenvectors of the adjoint matrix $H^{\dag}$
\begin{equation}
H^{\dag}= \left(
\begin{array}{cccccccccc}
V_{1}^{*} & \rho_{1}^{*} & 0 & 0 & 0 & ... & 0 & 0 & 0 & 0 \\
t_1^* & V_2^* & \rho_2^* & 0 & 0 & ... & 0 & 0 & 0 & 0 \\
0 & t_2^* & V_3^* & \rho_3^* & 0 & ... & 0 & 0 & 0 & 0 \\
0 & 0 & t_3^* & V_4^* & \rho_4^* & ... & 0 & 0 & 0 & 0 \\
... & ... & ... & .. & ... & ... & .. & ... & ... & ... \\
0 & 0 & 0 & 0 & 0 & ... & t_{L-3} ^{*}& V_{L-2}^* & \rho_{L-1}^{*} & 0 \\
0 & 0 & 0 & 0 & 0 & ... & 0 &  t_{L-2}^* & V_{L-1}^* & \rho_{L-1}^* \\
0 & 0 & 0 & 0 & 0 & ... & 0 &  0 &  t_{L-1}^* & V_{L}^*  \\
\end{array}
\right)
\end{equation}
corresponding to the eigenenergy $E^{*}_{\alpha}$. $\Psi_n^{(\alpha)}$ are often referred to as the left eigenvectors of $H$, while $\psi_n^{(\alpha)}$ are the right eigenvectosr of $H$. The following orthonormal condition between left and right eigenvectors holds
\begin{equation}
\langle \Psi^{(\alpha)}| \psi^{(\beta)} \rangle \equiv \sum_{n=1}^L \Psi_{n}^{(\alpha) *} \psi_n^{(\beta)}= \delta_{\alpha, \beta}.
\end{equation}
The Green function (resolvent) of $H$ 
\begin{equation}
G(E)=(E-H)^{-1}
\end{equation}
is a meromorphic function of $E$ with poles at $E=E_{\alpha}$. In fact, the following spectral representation of $G(E)$ holds
\begin{equation}
G_{n,m}(E)= \sum_{\alpha=1}^{L} \frac{\psi^{(\alpha ) }_n \Psi^{(\alpha) *}_m }{E-E_{\alpha}}
\end{equation}
which readily follows from the resolution of the identity $\sum_{\alpha} |\psi^{(\alpha)} \rangle \langle \Psi^{(\alpha)}|=\mathcal{I}$.
Equation (C6) shows that, for a simple eigenvalue $E_{\alpha}$, the residue of the pole of $G_{n,m}(E)$ at $E=E_{\alpha}$ is equal to 
\begin{equation}
\frac{1}{2 \pi i} \oint_{|E-E_{\alpha}|=0^+} dE \; G_{n,m}(E)
=\psi^{(\alpha )}_n \Psi^{(\alpha) *}_m.
\end{equation}
Let us now focus our attention to the element $G_{1,L}(E)$ of the Green function $(E-H)^{-1}$. From the definition of the inverse of a matrix, it follows that such an element is given by
\begin{equation}
G_{1,L}(E)=\frac{ \left\{ {\rm cofactor}(E-H) \right\}_{L,1}}{\det (E-H)} 
\end{equation}
where
\begin{widetext}
\begin{equation}
\left\{ {\rm cofactor} (E-H) \right\}_{L,1}  =
 (-1)^{L+1}   \left|
\begin{array}{cccccccc}
- t_1 & 0 & 0 & 0 &  ... & 0 & 0 & 0 \\
E-V_2 & - t_2 & 0 & ... & 0 & 0 & 0 \\
-\rho_2 &E-V_3 & -t_3 & ... & 0 & 0 & 0 \\
... & ... & ... & ... & ... & ... & ... & ... \\
0 & 0 & 0 & 0 & ... & E-V_{L-2} & -t_{L-2} & 0 \\
0 & 0 & 0 & 0 & ... & 0 & E-V_{L-1} & -t_{L-1}
\end{array}
\right|= \prod_{k=1}^{L-1} t_k
\end{equation}
\end{widetext}
is the cofactor of the element $(L,1)$ of the matrix $(E-H)$, and
\begin{equation}
\det (E-H) = \prod_{\alpha=1}^{L}(E-E_{\alpha})
\end{equation}
is the determinant of the matrix $(E-H)$. Substitution of Eqs.(C9) and (C10) into Eq.(C8) yields
\begin{equation}
G_{1,L}(E)=\frac{\prod_{k=1}^{L-1} t_k }{\prod_{\alpha=1}^{n} (E-E_{\alpha})}.
\end{equation}
From Eq.(C11) it follows that the residue of $G_{1,L}(E)$ at a simple eigenvalue $E_{\alpha}$ is given by
\begin{equation}
\frac{1}{2 \pi i} \oint_{|E-E_{\alpha}|=0^+} dE \; G_{1,L}(E)=\frac{\prod_{k=1}^{L-1} t_k }{\prod_{\beta=1, \; \beta \neq \alpha}^{n} (E-E_{\beta})}.
\end{equation}
A comparison of Eqs.(C7) and (C12) yields
\begin{equation}
\psi^{(\alpha )}_1 \Psi^{(\alpha) *}_L= \frac{\prod_{k=1}^{L-1} t_k }{\prod_{\beta=1, \; \beta \neq \alpha}^{n} (E-E_{\beta})}.
\end{equation}
To generalize the Thouless result relating the density of states and localization length \cite{R50}, we make the key assumption that the hopping amplitudes are symmetric and complex conjugate, i.e. 
\begin{equation}
\rho_n=t_n^*,
\end{equation}
 while the on-site potential $V_n$ can be complex valued. Such an assumption ensures that the following simple relation exists between left and right eigenvectors of $H$
 \begin{equation}
 \Psi^{(\alpha)}_n= \psi^{(\alpha) *}_n
 \end{equation}
so that Eq.(C13) takes the form
\begin{equation}
\psi^{(\alpha )}_1 \psi^{(\alpha) }_L= \frac{\prod_{k=1}^{L-1} t_k }{\prod_{\beta=1, \; \beta \neq \alpha}^{L} (E-E_{\beta})}.
\end{equation}
We now take the thermodynamic limit $L \rightarrow \infty$ and calculate the Lyapunov exponent $\lambda(E)$, i.e. the inverse of the localization length, for the eingenstate $\psi_n^{(\alpha)}$ of energy $E=E_{\alpha}$
\begin{equation}
\lambda(E)= -\lim_{L \rightarrow \infty} \frac{1}{L} \log \left| \frac{\psi_L^{(\alpha )}}{\psi_1^{(\alpha)}} \right|
\end{equation}
with $\lambda>0$ for localization.
Substitution of Eq.(C16) into Eq.(C17) yields
\begin{equation}
\lambda(E)= \lim_{L \rightarrow \infty} \frac{1}{L} \sum_{\beta=1 \;, \beta \neq \alpha}^{L} \log |E-E_{\beta}|-  \lim_{L \rightarrow \infty}  \frac{1}{L} \sum_{k=1}^{L-1} \log |t_k|
\end{equation}
For the Hamiltonian $\hat{H}$ describing the non-Hermitian AAH model, the hopping amplitudes are equal and real, i.e. $t_k=J$ independent of $k$, so that Eq.(C18) takes the form of Eq.(28) given in the main text. 
More generally, after introduction of the density of states in complex energy plane, in the thermodynamic limit $L \rightarrow \infty$ a relation between localization length and an integral of the density of states in complex energy plane can be established, which is given by Eq.(29) in the main text.

\end{document}